\documentclass[aps,prl,twocolumn,groupedaddress,superscriptaddress]{revtex4-1}
\usepackage{color}
\usepackage{amsmath}
\usepackage{amssymb}
\usepackage{graphicx}
\usepackage{epstopdf}
\usepackage[hidelinks,colorlinks=true,linkcolor=blue,citecolor=blue,anchorcolor=black,filecolor=cyan,menucolor=red,runcolor=cyan,urlcolor=magenta]{hyperref}
\usepackage{bm}
\usepackage{color}

\makeatother

\begin{document}

\title{Measuring the topological phase transition via the single-particle density matrix}
\author{Jun-Hui Zheng}
\affiliation{Institut f{\"u}r Theoretische Physik, Goethe-Universit{\"a}t, 60438
Frankfurt am Main, Germany}
\author{Bernhard Irsigler}
\affiliation{Institut f{\"u}r Theoretische Physik, Goethe-Universit{\"a}t, 60438
Frankfurt am Main, Germany}
\author{Lijia Jiang}
\affiliation{Frankfurt Institute for Advanced Studies, 60438 Frankfurt am Main, Germany}
\author{Christof Weitenberg}
\affiliation{Institut f{\"u}r Laserphysik, Universit{\"a}t Hamburg, Luruper Chaussee 149, 22761 Hamburg, Germany}
\affiliation{Hamburg Centre for Ultrafast Imaging, Luruper Chaussee 149, 22761 Hamburg, Germany}
\author{Walter Hofstetter}
\affiliation{Institut f{\"u}r Theoretische Physik, Goethe-Universit{\"a}t, 60438
Frankfurt am Main, Germany}

\date{\today}

\begin{abstract}

We discuss the topological phase transition of the  spin-$\frac{1}{2}$ fermionic Haldane model with repulsive on-site interaction. 
We show that the Berry curvature of the topological Hamiltonian, the first Chern number, and the topological phase transition point can be extracted from the single-particle density matrix for this interacting system. Furthermore, we design a tomography scheme for the single-particle density matrix of interacting fermionic two-band models in experimental realizations with cold atoms in optical lattices.
\end{abstract}
\maketitle

Topological insulators are a fascinating new phase without a local order parameter  \cite{Hassan2010rmp,Xiao2010rmp}. They  have been observed in solid-state materials  \cite{Gehring2013nl}, but have also been realized in quantum simulators such as photonic wave\-guides \cite{Rechtsman2013prl}  and ultracold atoms \cite{Aidelsburger2013prl,Jotzu2014Nat,Flaschner2016Sci}. In two-dimensional systems, topology can be captured by the Chern number (ChN) as the topological index, which is given by the sum of the integral of the Berry curvature in the Brillouin zone over all occupied bands  \cite{Xiao2010rmp}.  Topological insulators, which are characterized by a non-zero Chern number, possess robust conducting edge states at their boundaries. The number of edge states is equal to the Chern number (ChN) for noninteracting systems, according to the bulk-edge correspondence  \cite{Hassan2010rmp}. In solid-state systems and photonic systems, the topology is often revealed via the edge states  \cite{Hassan2010rmp,Rechtsman2013prl}, while in quantum gas experiments, also the Berry curvature can be reconstructed from quench dynamics \cite{Duca2015Sci, Flaschner2016Sci}.


Generalized to interacting systems, the ChN is expressed by the Ishikawa-Matsuyama formula in terms of the  single-particle Green's function \cite{Ishikawa1986}. It still reflects the number of  quasiparticle edge states when the interaction is weak or moderate, even though the bulk-edge correspondence breaks down in some situations with strong interactions \cite{Gurarie2011prb,You2014prb,He2016prb}. On the other hand, it was proven that the ChN can be evaluated by mapping to a noninteracting topological Hamiltonian determined by the zero-frequency Green's function, $H_{t}\equiv - 1/ G_{{\bf k}, i\omega =0}$ \cite{Wang2012prx}, or via the quasiparticle Berry curvature \cite{Shindou2008,Shindou2008prb,Zheng2018,Wong2013,Sengupta2015}.  Numerical simulations confirm that interaction could induce topologically nontrivial phases for specific systems \cite{Abanin2012prl,Kumar2016prb,Hofstetter2018,Rachel2018,Irsigler2018,Zheng2018a}.
However, these conclusions have so far not been confirmed experimentally. The main reason is that it is still unclear which observables correspond to the topological Hamiltonian and the quasiparticle Berry curvature.

In this letter, we consider the half-filled two-band model in a bipartite lattice with repulsive interaction.
We illustrate that the Berry curvature of the topological Hamiltonian, the first Chern number, and the phase transition point can be extracted from the single-particle density matrix (SPDM) of the interacting system. The elements of the SPDM are $\rho_{{\bf k},\alpha\beta} =\langle \hat{c}^\dagger_{{\bf k}\alpha}   \hat{c}_{{\bf k}\beta} \rangle$, where $ \hat{c}^\dagger_{{\bf k}\alpha}$ and $ \hat{c}_{{\bf k}\beta}$ are the  fermionic creation and annihilation operators with momentum $\bf k$, and $\alpha,\beta$ represent the pseudospin from A-B sublattice. Furthermore, we develop a scheme of tomography for the SPDM of interacting fermions in two-dimensional optical lattices with a two-sublattice structure. This scheme involves time-of-flight (TOF) imaging of the momentum distribution following different sudden quenches, which can be implemented in cold atom experiments. Our method generalizes the scheme of tomographic measurement of pure or mixed states proposed in Refs.\,\cite{Hauke2014prl, Ardila2018ar,Tarnowski2017prl} and realized in Ref.\,\cite{Flaschner2016Sci}.

The topological Hamiltonian carries the full information on the topology of the interacting system and is theoretically important for understanding the topological phase transition via analogy with the noninteracting system \citep{Wang2012prx}, yet it is not a physical observable. The following statement builds a link between the topological Hamiltonian and the SPDM for  half-filled fermionic two-band systems, which paves the way to probe it experimentally:

If the intrinsic quasiparticle linewidths $\gamma_{p}({\bf k})$ and $\gamma_{h}({\bf k})$  are much smaller than the quasiparticle energy $[\epsilon_{p}({\bf k})>0]$ and the quasihole energy $[\epsilon_{h}({\bf k})<0]$ respectively, i.e., $\gamma_{p}({\bf k})\ll |\epsilon_{p}({\bf k})|$ and $\gamma_{h}({\bf k})\ll |\epsilon_{h}({\bf k})|$, then the inverse of the topological Hamiltonian can be approximated as
\begin{equation} \label{green_density}
H_t^{-1}({\bf k}) \simeq   \frac{ \rho^T_{\bf k} }{\epsilon_h({\bf k})}
+ \frac{{\bf 1} -\rho^T_{\bf k}}{\epsilon_p({\bf k})} ,
\end{equation}
where  $\rho^T_{\bf k}$ is the transpose of the SPDM, $\bf 1$ is the $2\times 2$ identity matrix.

In order to prove this, we start from the Lehmann representation of the Green's function at zero temperature
\begin{equation}  \label{green_lehmann}
\mathcal{G}^{\alpha\beta}_{{\bf k}, i\omega} = \sum_{\eta} \Big[ \frac{\langle 0 | \hat{c}_{{\bf k} \alpha} | \eta \rangle\langle \eta| \hat{c}^\dagger_{{\bf k}\beta} |0\rangle }{i\omega - E_{ \eta}}+ \frac{\langle \bar{\eta} | \hat{c}_{{\bf k} \alpha} | 0\rangle\langle 0| \hat{c}^\dagger_{{\bf k}\beta} | \bar{\eta} \rangle }{i\omega + E_{\bar{\eta}} } \Big], 
\end{equation}
where $|0\rangle$ is the many-body ground state with zero energy.  ${\eta}$ and $\bar{\eta}$ refer to the  
excitations ($E_{\eta},E_{\bar{\eta}} >0$). For each given momentum, the spectral density is given by the imaginary part of the trace  of the retarded Green's function, $\varrho_{\bf k}(\omega) = \sum_{\eta\alpha} \big[ {\left|\langle 0 | \hat{c}_{{\bf k} \alpha} | \eta \rangle \right|^2}\delta({\omega - E_{ \eta}})+ { \left|\langle \bar{\eta} | \hat{c}_{{\bf k} \alpha} | 0\rangle \right|^2 }\delta({\omega + E_{\bar{\eta}} }) ]$.  The coefficient $|\langle 0 | \hat{c}_{{\bf k} \alpha} | \eta \rangle|^2 $ or $ |\langle \bar{\eta} | \hat{c}_{{\bf k} \alpha} | 0\rangle|^2$  becomes a nonnegligible contribution only when the energy of the many-body state is near to the quasiparticle energy, i.e., $\big| E_{\eta} - \epsilon_p \big| \lesssim \mathcal{O}[\gamma_p ] $ or $\big|E_{\bar{\eta}} + \epsilon_h \big| \lesssim \mathcal{O}[\gamma_h ]$. When the linewidth is rather small compared to the quasiparticle energy,  we have ${1}/{E_\eta} \simeq {1}/{\epsilon_p}$ and  ${1}/{E_{\bar\eta}} \simeq -{1}/{\epsilon_h}$ for the contribution to $\varrho_{\bf k}$ and $\mathcal{G}^{\alpha\beta}_{{\bf k},i\omega=0}$. By using 
$\langle 0| \hat{c}^\dagger_{{\bf k}\beta} \hat{c}_{{\bf k} \alpha} | 0\rangle= \rho_{{\bf k},\beta\alpha}$ and $ [\hat{c}_{{\bf k} \alpha}, \hat{c}^\dagger_{{\bf k}\beta}] = \delta_{\alpha\beta}$, we indeed obtain Eq.\,\eqref{green_density} from Eq.\,\eqref{green_lehmann} at zero frequency. The error for this approximation is of order $\gamma_{p(h)}/\epsilon_{p(h)}$.

Eq.\,\eqref{green_density} shows that 
$H_t({\bf k})$ and $\rho^T_{\bf k}$ have exactly the same eigenvectors and the lower band of the former is mapped onto the higher band of the latter. 
This allows us to obtain the Berry curvature of the topological Hamiltonian through measuring the SPDM. 
In addition, Eq.\,\eqref{green_density} still holds when the temperature $T$ is finite but much smaller than the gap. The additional error is suppressed exponentially by $\exp[{-|\epsilon_{p(h)}|/k_B T}]$.

\begin{figure}
\includegraphics[width=\columnwidth]{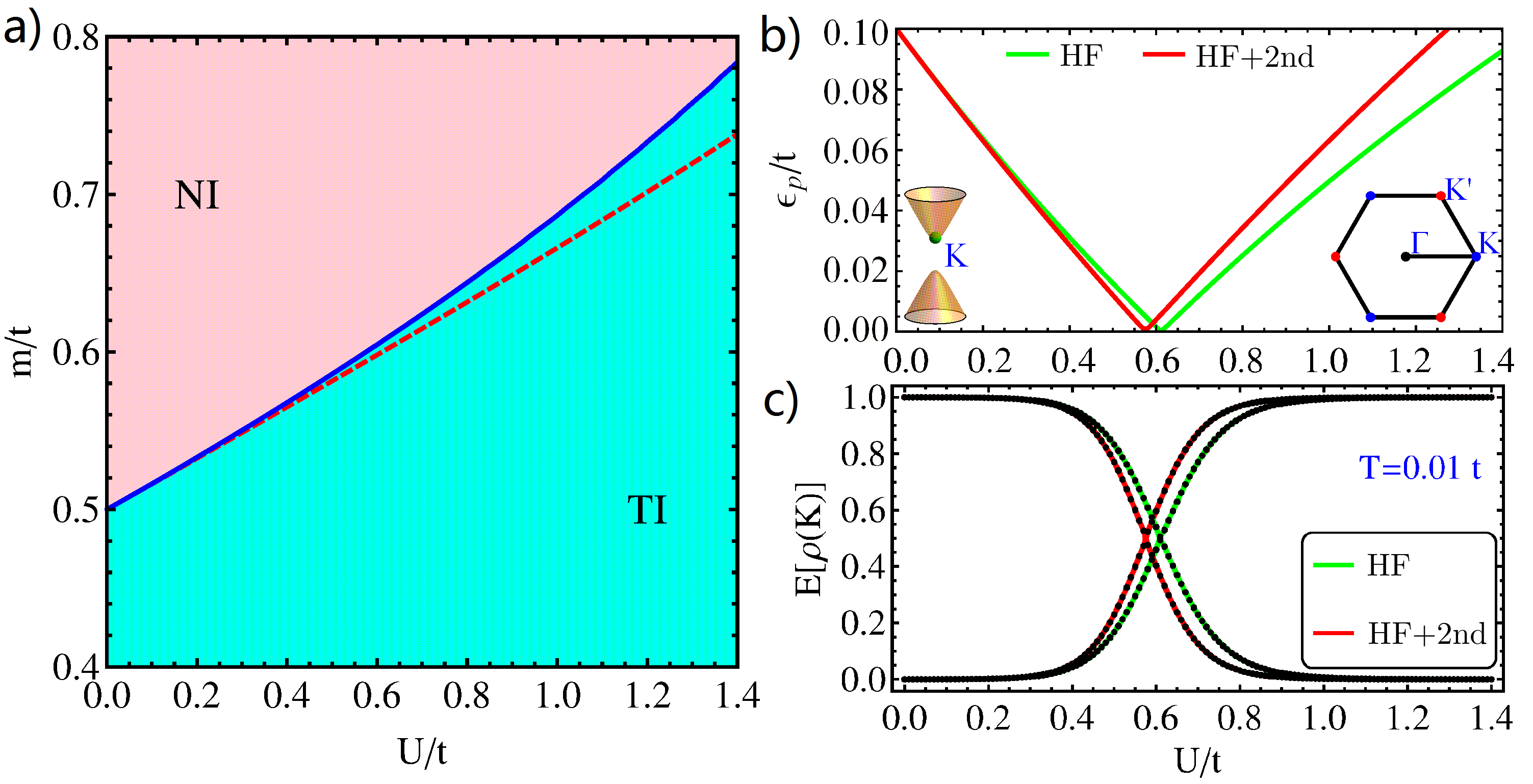}\centering
\caption{(a) The phase diagram of the Haldane model with different values of the staggered potential and the interaction strength. The red dashed and blue solid lines are given by HF and HF+2nd approximation, respectively.  (b) The quasiparticle energy of the upper band at the Dirac point (K), for HF and  HF+2nd approximation. (c) The eigenvalues of the SPDM at the K point for $T=0.01 t$. For (b) and (c), $m=0.6 t$.}
\label{PTD}
\end{figure}

Let us consider the spin-$\frac{1}{2}$ Haldane model in a hexagonal optical lattice, which has been realized as a Floquet system in cold atom experiments \cite{Jotzu2014Nat,Flaschner2016Sci}. The Hamiltonian reads
\begin{equation}
\hat{H}_0 = - t\sum_{\langle i j\rangle s} \hat{c}_{is}^{\dagger} \hat{c}_{js} +  \lambda  \sum_{\langle\langle i j\rangle\rangle s}e^{i\phi \nu_{ij}}\hat{c}_{is}^{\dagger}\hat{c}_{js}
 +m \sum_{is}\xi_i \hat{c}_{is}^{\dagger} \hat{c}_{is}, \label{haldane}
\end{equation}
where the first and second terms are the nearest and the next-nearest neighbor hopping terms. $s$ refers to spin $\uparrow$ and $\downarrow$. $\nu_{ij} = \pm 1$, which is related to the hopping path. In the following, we restrict ourselves to the case of $\phi=\pi/2$, which maximally breaks time reversal symmetry. The third term is a staggered potential with $\xi_i = 1$ for sublattice A and  $\xi_i = -1$ for sublattice B. 
The system displays a transition into a normal insulator from the quantum Hall phase when $|m|$ becomes larger than $3\sqrt{3}|\lambda|$. The energy gap of the system is $2|m -3\sqrt{3}\lambda|$ for $m,\lambda>0$. The on-site interaction reads
\begin{equation}
\hat{H}_{\text{int}} = U \sum_{i} \hat{n}_{i \uparrow}\hat{n}_{i\downarrow}.
\label{haldane2}
\end{equation}
The system has SU(2) symmetry in spin space. Note that an interacting Floquet system contains additional subtleties such as micromotion corrections to the interaction \cite{Anisimovas2015prb}. With our static effective model given by Eqs.\,\eqref{haldane} and \eqref{haldane2}, we focus on the high frequency regime, where these corrections are suppressed \cite{Eckardt2017rmp}. Related interaction effects in static models can be found in Refs.\,\cite{Imriska2016prb,Wu2016prb,Vanhala2016prl,Rubio-Garcia2018njp}.

For weak interaction, using the Hartree-Fock (HF) approximation and HF plus the  second-order perturbation correction (HF+2nd), respectively, we plot the phase diagram in Fig.\,\ref{PTD}a for the case  $3\sqrt{3}\lambda = 0.5 t$.  The HF approximation yields a renormalized staggered potential, $m \rightarrow m+ \frac{U}{2}\big[\langle \hat{n}_{\text{A}\downarrow} \rangle-\langle \hat{n}_{\text{B} \downarrow} \rangle\big]$, where $\hat{n}_{\text{A(B)} \downarrow}$ is the number operator of spin down at each site of sublattice A(B). The repulsive interaction effectively smoothens the staggered potential, and induces the  topological insulator phase, which is consistent with the result shown in Ref.\,\cite{Vanhala2016prl}.  For $m= 0.6 t$, we show the quasiparticle energy at the Dirac point K within HF and HF+2nd approximation in Fig.\,\ref{PTD}b. The gap of the system is exactly twice this energy due to particle-hole symmetry. The interaction closes the gap and inverts the bands  at $U \simeq 0.6 t$.

\begin{figure}
\includegraphics[width=0.95\columnwidth]{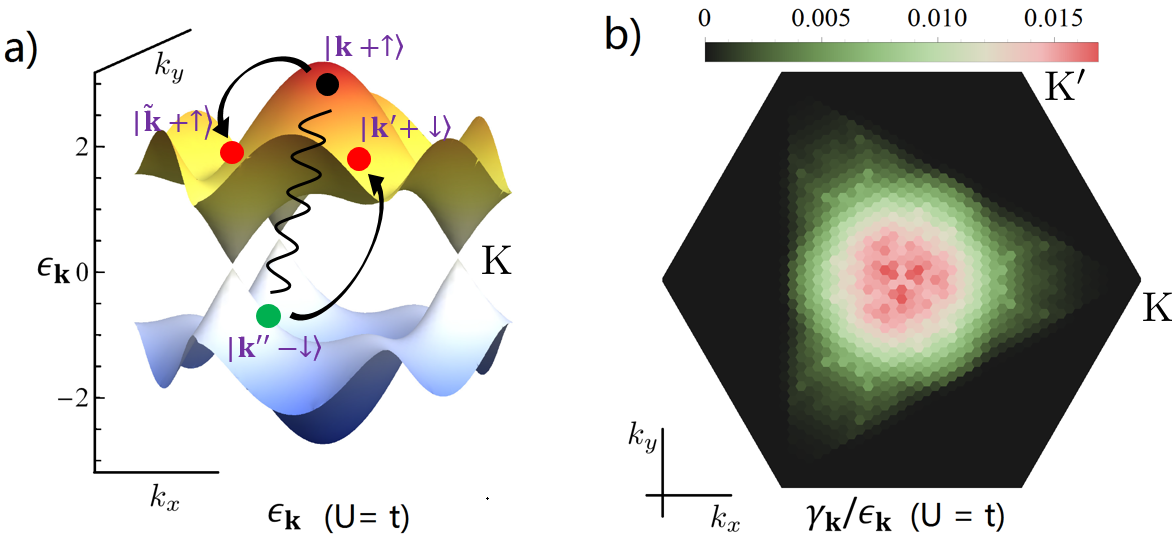}
\caption{(a) The HF quasiparticle spectrum and the sketch of the collision between two HF states, and (b) the ratio of the linewidth to the quasiparticle energy. The region where the linewidth vanishes is due to energy and momentum conservation.}
\label{liferatio}
\end{figure}

In the following, we confirm that the linewidth $\gamma_{\bf k}$ is rather small for the weak interaction regime. The linewidth of a HF quasiparticle excitation (corresponding to the HF approximation) can be obtained by considering all collision channels with one particle from the lower band (see Fig.\,\ref{liferatio}a). Using Fermi's Golden Rule, we obtain the linewidth for the quasiparticle state $|{\bf k} + \uparrow \rangle$, 
\begin{eqnarray}
\gamma_{\bf k} &=& {2 \pi} \times  \frac{ A^2_{r}}{(2\pi)^4} \int d^2 {\bf k}' d^2{\bf k}'' \delta(\epsilon_{\bf k}-\epsilon_{\bf k'}+ \epsilon_{\bf k''}-\epsilon_{\bf \tilde{k}}) \notag\\
&& \times \left|\big\langle {\bf \tilde{k}  }+ \uparrow, {\bf k' }+ \downarrow\big| \hat{H}_\text{int}\big|{\bf k} + \uparrow, {\bf k''} - \downarrow\big \rangle\right|^2 , \label{fermig}
\end{eqnarray}
 where $A_{r}$ is the area of the system and $|{\bf k} \pm \uparrow \rangle$ is the eigenstate of the higher (lower) band with spin up within the HF approximation.  Each energy level is two-fold spin degenerate. The two outgoing particles occupy states in the higher band, since the lower band is filled. Momentum conservation demands $\bf \tilde{k} = \bf k'' +\bf k-\bf k'$. The $\delta$-function in Eq.\,\eqref{fermig} stems from energy conservation. The phase space of the final states is constrained by momentum and energy conservation. In particular, for the quasiparticle  at the Dirac point K the linewidth vanishes for zero temperature, since all collision channels are forbidden. 
In comparison, a quasiparticle excitation with a higher energy has a larger linewidth and ratio $\gamma_{\bf k}/\epsilon_{\bf k}$ due to a larger phase space of the final states (see Fig.\,\ref{liferatio}b). 
The linewidth as a function of interaction strength, Eq.\,\eqref{fermig}, can formally be parameterized as 
 $c_1 U^2 (1+ c_2 U + \cdots)$, where the first $U^2$ directly arises from $\hat{H}_{\text{int}}$ and the part $c_2 U$ is due to the interaction dependent HF states. For weak interaction, the linewidth increases quadratically as a function of the interaction strength.  In Fig.\,\ref{liferatio}, we show the HF quasiparticle energy and the ratio of the linewidth to the energy for $ U=t$. 
For different interaction strengths, the linewidth has a similar profile in momentum space but with an interaction-dependent rescaling. A large interaction enhances the linewidth, and thus the ratio $\gamma_{\bf k}/\epsilon_{\bf k}$. 
We find that up to $U= t$, the linewidth is still rather small compared to the energy ($< 2.6\%$) for $m \in [0, t]$. A similar conclusion  can  be drawn for quasihole states. This confirms the validity of the approximation  \eqref{green_density}.

The ratio $\gamma_{\bf k}/\epsilon_{\bf k}$ also reflects how much the quasiparticle differs from a single-particle pure state. In principle, when the interaction becomes stronger, the deviation 
 increases. On the other hand, also the temperature $T$ can mix states. For $T=0.01 t$, we plot the eigenvalues of the SPDM within HF and HF+2nd approximation, respectively, for the K point in Fig.\,\ref{PTD}c. The position of the gap closing of   the SPDM   almost  coincides with that of the energy in Fig.\,\ref{PTD}b. This means that the topological phase transition point can be obtained from the gap closing point of the SPDM as expected.  The small deviation from the real phase  transition point  stems from finite $T$ and linewidth at the $\Gamma$ point, respectively.

We have shown that the higher band of $\rho_{\bf k}^T$ provides information on topological properties of the lower energy band of the system. In the following, we illustrate how to measure it in cold atom experiments. Including finite temperature and interaction effects, the many-body density matrix  of an interacting system is $\mathcal{P}_M = \sum_\eta p_\eta | \eta \rangle \langle \eta|$, where $| \eta \rangle$ is a many-body energy eigenstate and $p_\eta$ is the thermal equilibrium probability distribution function with the constraint $\sum_\eta p_\eta =1$. The SPDM becomes ${\rho}_{{\bf k},\alpha\beta} = \text{Tr} [\hat{c}^\dagger_{{\bf k}\alpha}   \hat{c}_{ {\bf k}\beta}  \mathcal{P}_M]$.  Here and in the following, the spin index is dropped due to SU(2) symmetry. $\alpha$ and $\beta$ are the pseudospin sublattice index (A, B). The transpose of the SPDM  can be represented as
\begin{equation} \label{pauli}
\rho^T_{\bf k} = \frac{1}{2} \sum_{i=0}^3 a_{{\bf k},i} \sigma_i,
\end{equation}
where $\sigma_{1(2,3)} $ is the Pauli matrix and $\sigma_0 = \bf 1$. The coefficients are $a_{{\bf k},i}= {\rm Tr}[\rho^T_{\bf{k}} \sigma_i] =\sum_\eta p_\eta  \langle \eta|  \hat{c}^\dagger_{\bf{k}} \sigma_i\hat{c}_{\bf{k}}| \eta \rangle $,
where  $\hat{c}_{\bf{k}} = (\hat{c}_{{\bf k}\text{A}},\hat{c}_{{\bf k}\text{B}})^T$.  Note that $ a_{{\bf k},0} >0 $ is the total density $\rho_{{\bf k}, \text{AA}}+ \rho_{{\bf k}, \text{BB}}$, and it equals 1 for the half filling case with particle-hole symmetry. 

Quench dynamics can be used to reconstruct the vector ${\bm a}_{\bf k} \equiv (a_{{\bf k},1}, a_{{\bf k},2}, a_{{\bf k},3})$. Let us suppose that the system is suddenly quenched to a new noninteracting Hamiltonian $\hat{\mathcal{H}} = \frac{\Omega}{2} \sum_{\bf k} \hat{c}^\dagger_{\bf{k}}{\bm\sigma\cdot}{{\bm d}_{\bf k}}  \hat{c}_{\bf{k}}$ at the time $\tau=0$, where ${{\bm d}_{\bf k}}$ is a momentum-dependent unit vector.  The coefficients of $\rho^T_{\bf k}(\tau)$ become
$ a_{{\bf{k}}, j}(\tau) =  \sum_\eta p_\eta  \langle  \eta|   \hat{c}^\dagger_{\bf{k}}  V_{\bf k}^\dagger(\tau)\sigma_j {V_{\bf k}}(\tau) \hat{c}_{\bf{k}} |  \eta \rangle$ after evolution to time $\tau>0$, where ${V_{\bf k}}(\tau) = e^{-i \tau (\Omega/2) {\bm\sigma\cdot}{{\bm d}_{\bf k}}   }$ is a $2\times 2$ matrix. Since $ V_{\bf k}^\dagger(\tau)\sigma_j {V_{\bf k}}(\tau) $ is a linear combination of $\sigma_i$, this formula links the coefficients at time $\tau$ to those at time $\tau=0$. The evolution effectively rotates the vector ${\bm a}_{\bf k}$, and $a_{{\bf{k}},0}$ is time-independent after the quench. Thus, the initial SPDM can be deduced from the final coefficients. However, in TOF experiments, not all of the  final coefficients can be recorded. The density operator of particles in momentum space observed in TOF experiments is $
\hat{N}_{\text{TOF}}({\bf k}) \simeq  |\tilde{w}({\bf k})|^2 \sum_{\bf{ R R^\prime}} e^{-i {\bf{k \cdot (R-R^\prime)}}}  \hat{c}^\dagger_{{\bf R^\prime}}\hat{c}_{\bf R}$,
where $\bf R$ and $\bf R'$ denote lattice sites and $\tilde{w}(\bf k)$ is the Fourier transformation of the Wannier function \cite{Bloch2008rmp}. So the particle density observed is
\begin{eqnarray}
{N}_{\text{TOF}}({\bf k}) &\simeq & |\tilde{w}({\bf k})|^2 \langle (\hat{c}^\dagger_{{\bf k} \text{A}} + \hat{c}^\dagger_{{\bf k } \text{B}})(\hat{c}_{{\bf k} \text{A}} + \hat{c}_{{\bf k} \text{B}}) \rangle
  \notag \\
& = & |\tilde{w}({\bf k})|^2 [a_{{\bf{k}},0}^F+ a_{{\bf k},1}^F],
\end{eqnarray}
where $a_{{\bf k},0}^F = a_{{\bf k},0} = 1$ and  $a^F_{{\bf k},1}$ are the components of the final SPDM  at the time before the free expansion. Only the component $a_{{\bf k},1}^F$ can be detected.

Through rotating the initial vector ${\bm a}_{\bf k}$  during the dynamics after the quench, we can reconstruct ${\bm a}_{\bf k}$ by detecting its projection onto the first component.
In the first protocol, the system is suddenly quenched  to $\hat{\mathcal{H}}$ with $\bm{d}_{\bf k} = (0,0,1)$ at the time $\tau=0$, which can be realized by switching off all tunneling and interaction but with the staggered potential ${\Omega}/{2}$ remaining  \cite{Hauke2014prl}. The rotation couples $a_{{\bf k},1}$ and $a_{{\bf k},2}$, and we have  $a_{{\bf{k}}, 1}(\tau) =  \cos(\Omega \tau) a_{{\bf{k}}, 1} - \sin(\Omega \tau) a_{{\bf{k}}, 2}$. If the atoms are completely released at time $\tau$, then the particle density observed by TOF imaging is 
\begin{equation}
{N}^{\text{I}}_{\text{TOF}}({\bf k}, \tau) 
\propto
1 +  \cos(\Omega \tau) a_{{\bf{k}}, 1} - \sin(\Omega \tau) a_{{\bf{k}}, 2}.  \label{den}
\end{equation}
The protocol is the same as that for a single-particle pure state (SPPS) in noninteracting systems  \cite{Hauke2014prl}. By fitting to the experimental data,  both $a_{{\bf{k}}, 1}$ and $a_{{\bf{k}}, 2}$ can be obtained from the oscillating behavior of ${N}^{\text{I}}_{\text{TOF}}({\bf k}, \tau) $.  For the SPPS, $|{\bm a}_{{\bf k}}| =1$ so that $|{ a}_{{\bf k},3}|$ can be obtained from the known $a_{{\bf{k}}, 1}$ and $a_{{\bf{k}},2}$. This is not true for  a general density matrix where $|{\bm a}_{{\bf k}}| \leq 1$. Additional experiments involving components  $a_{{\bf{k}}, 1}$ and $a_{{\bf{k}},3}$ are needed for detecting $a_{{\bf{k}},3}$.

\begin{figure}
\includegraphics[width=0.9\columnwidth]{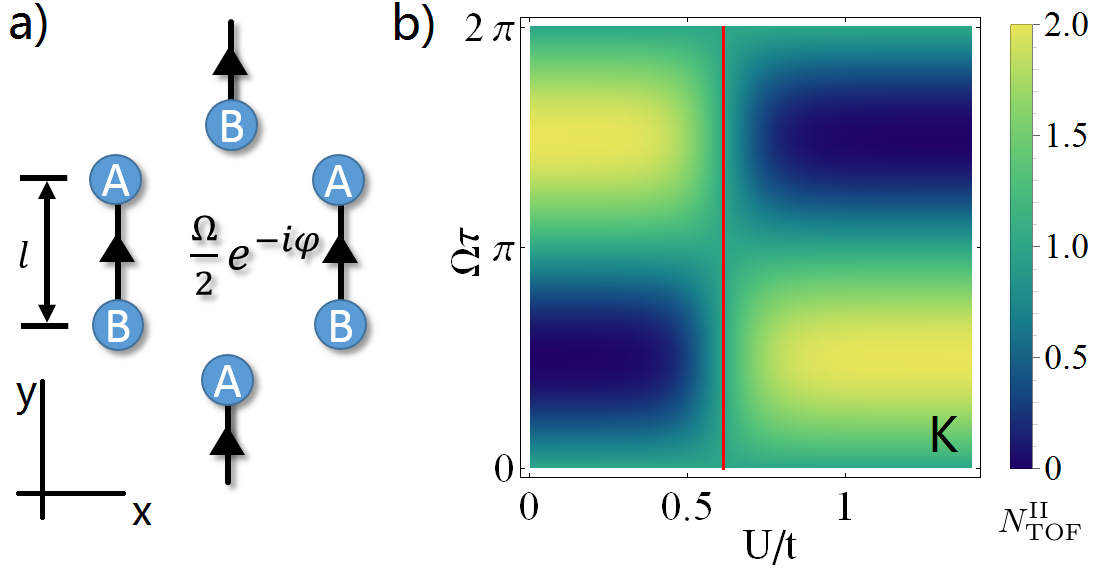}
\caption{(a) Second quench protocol and (b) particle density oscillation observed in TOF experiment for the Dirac point ${\bf k} = (4\pi/3\sqrt{3}l,0)$, by using the second protocol with $\varphi = \pi/2$. The red line represents the phase transition. T=0.01t.}
\label{quench}
\end{figure}

The second protocol uses the quench channel $\hat{\mathcal{H}}$ with  ${\bm d}({\bf k}) = (\cos(k_y l+\varphi), \sin(k_y l+\varphi), 0 )$, where $l$ is the lattice constant. This Hamiltonian can be realized by  switching on (laser assisted) tunneling ${\Omega e^{-i\varphi}}/{2}$ between A-B sublattices only along $y$-direction, as shown in Fig.\,\ref{quench}a. This Hamiltonian induces a similar precession dynamics on the Bloch sphere as the first protocol, but now along a vector, which lies in the xy-plane. The coefficient then becomes 
\begin{eqnarray}
a_{{\bf{k}}, 1}(\tau) &=&  [\cos^2(\Omega \tau/2) + \sin^2(\Omega \tau/2)\cos(2k_y l + 2 \varphi)] a_{{\bf{k}}, 1}  \notag \\
&& +  \sin^2(\Omega \tau/2)\sin(2k_y l + 2 \varphi) a_{{\bf{k}}, 2} \notag \\
&&
+\sin(\Omega  \tau)\sin(k_y l +\varphi) a_{{\bf{k}} ,3}.
\end{eqnarray}

Using known $a_{{\bf{k}} ,1}$ and $a_{{\bf{k}}, 2}$, we can get $a_{{\bf{k}},3}$ by detecting the  particle density ${N}^{\text{II}}_{\text{TOF}}({\bf k},\tau) \propto 1+ a_{{\bf{k}} ,1}(\tau)$, except for the points $\bf k$ with $\sin(k_y l +\varphi) =0$. At these points, ${\bm d}({\bf k}) = (1,0,0)$, which cannot generate an effective rotation that couples $a_{{\bf{k}}, 1}$ and $a_{{\bf{k}},3}$. To obtain $a_{{\bf{k}} ,3}$ in the whole Brillouin zone, a similar  experiment with $\varphi \rightarrow \varphi + \pi/2$ can be implemented.  One can choose the two 
experiments with $\varphi =0$ (normal tunneling) and $\varphi =\pi/2$ (laser assisted tunneling), respectively.
Since the second protocol directly accesses $a_{{\bf{k}},3}$, there is no missing information on northern or southern hemisphere as it appears for the quench on flat bands \cite{Hauke2014prl}.

Note that all the different Hamiltonians discussed above could be realized by starting with a static lattice with large AB-offset and shaking \cite{Flaschner2016Sci,Weitenberg2017ar}. Specifically, circular shaking is used for simulating the Haldane model, while asymmetric linear shaking along the y-direction can be used for realizing the situation in Fig.\,\ref{quench}a \cite{Struck2012prl}. For realizing the first protocol, the quench can simply be realized by switching off the shaking, which was used to realize the Haldane model before the quench. The interaction can be switched off by using a Feshbach resonance \cite{Chin2010rmp} or by tuning the confinement strength along $z$-direction for a transverse confinement optical lattice. 
The time scale for ramping the interaction to zero should be much smaller than the time scale $1/\Omega$ and $1/U$, so that interaction effects during quench dynamics can be omitted.

\begin{figure}
\includegraphics[width=1.0\columnwidth]{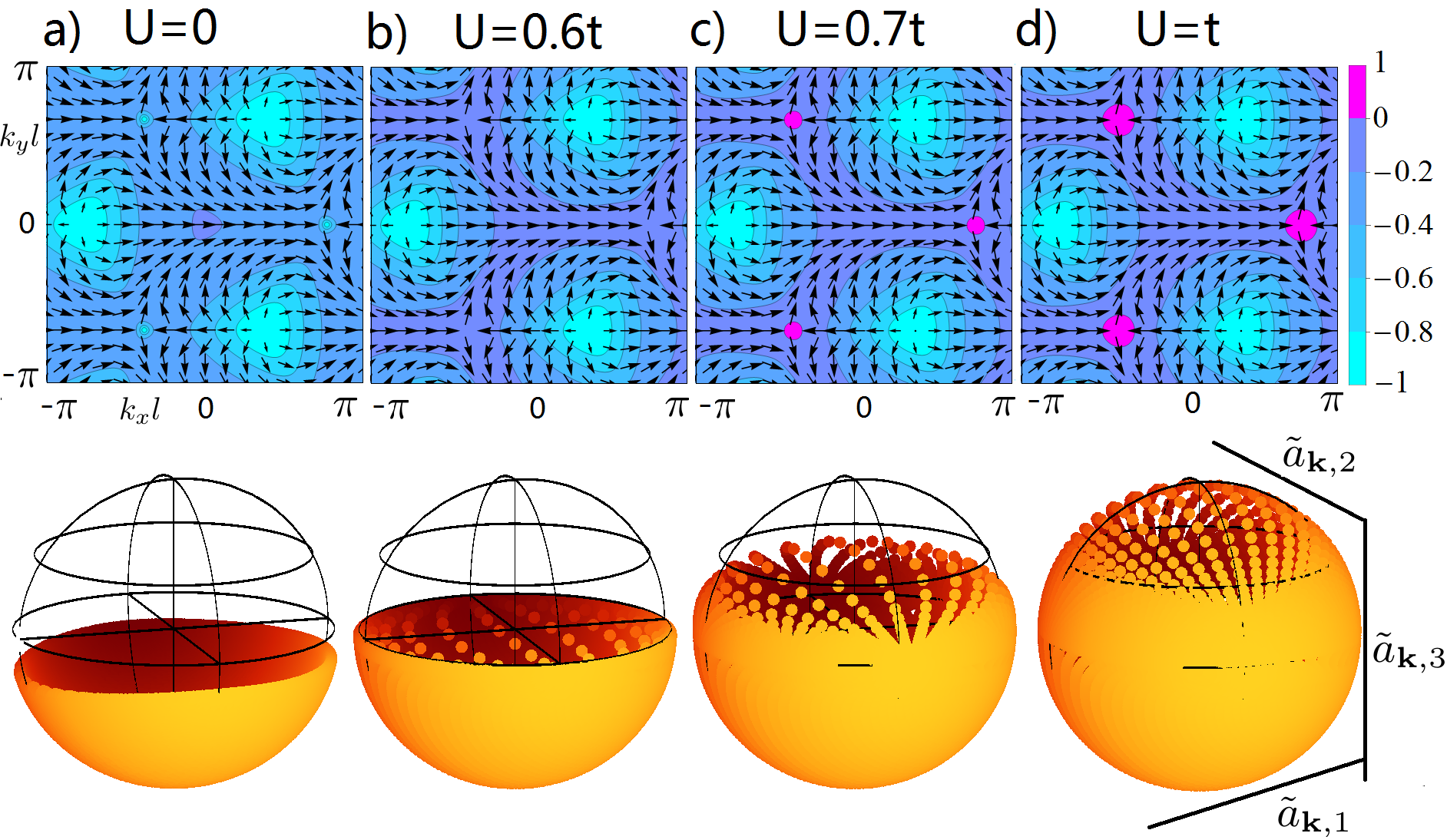}
\caption{Upper: vector plot of ($\tilde{a}_{{\bf k},1},\tilde{a}_{{\bf k},2}$) and density plot of $\tilde{a}_{{\bf k},3}$. Bottom: the 3D plot of $\bm{\tilde{a}}_{{\bf k}}$. T =0.01t.}
\label{berryc}
\end{figure}

The Berry curvature can then be extracted from the known ${\bm a}_{{\bf k}}$. Note that in the  Fourier transformed basis,  $\hat{c}_{{\bf k}\text{A(B)}} \propto \sum_{{\bf R} \in \text{A(B)}} \hat{c}_{\bf R} e^{-i{\bf k}\cdot {\bf R}}$, the Hamiltonian is not periodic but with an additional unitary transformation after translating by a reciprocal lattice vector $\bf Q$. We obtain $h({\bf k + Q}) = U^\dagger_{\bf \bf Q} h({\bf k}) U_{\bf Q} $, where $U_{\bf Q} =\text{diag}(1,  e^{-i Q_y l})$ and $h({\bf k})$ is the matrix representation of the noninteracting Hamiltonian $\hat{H}_0$. Thus we  introduce the unitary transformation $(\hat{\tilde {c}}_{{\bf k}\text{A}}, \hat{\tilde {c}}_{{\bf k}\text{B}}) = (\hat{c}_{{\bf k}\text{A}}, \hat{c}_{{\bf k}\text{B}} e^{-i k_y l}) $ to render the Hamiltonian periodic. The components for the periodic SPDM $\langle 
\hat{\tilde {c}}^{\dagger}_{{\bf k}\alpha} \hat{\tilde {c}}_{{\bf k}\beta} \rangle$ are
${ \tilde{\bm a}}_{{\bf k}} = (\cos(k_y l)a_{{\bf k},1}+ \sin(k_y l) a_{{\bf k},2}, \cos(k_y l) a_{{\bf k},2}-\sin(k_y l)a_{{\bf k},1}, a_{{\bf k},3})$.
We plot the result of ${\bm \tilde{a}}_{\bf k}$ for different interaction strengths in Fig.\,\ref{berryc}. The two-dimensional vector $({\tilde{ a}}_{{\bf k},1},  {\tilde{ a}}_{{\bf k},2})$ has an opposite winding behavior circuiting the Dirac points K and K$'$, as in noninteracting systems \cite{Hauke2014prl}. The third component ${\tilde{ a}}_{{\bf k},3}$ for the K point moves from the south  pole to the north pole when increasing the interaction strength. It changes its sign when $U\simeq 0.6t$.   The vector  ${\bm \tilde{a}}_{\bf k}$ maps the Brillouin zone to a   closed curved  surface in  three-dimensional space.  For the noninteracting case, it looks like a deflated ball. Interaction inflates this ball to be round.  The condition for a topological phase of the higher (and lower) band of $\rho^T_{\bf k}$ is that the origin is enclosed by that surface \cite{Niu2010rmp}.  This coincides with whether  $a_{{\bf k},3}$ at the K point is positive. Recall that for a single-particle pure state $|{\bm a}_{\bf k}|=1$ and it lives on the surface of the sphere (see Fig.\,\ref{berryc}a). With interaction and finite temperature, ${\bm a}_{\bf k}$ can lie within the sphere and the topological phase transition occurs mildly.  The Berry curvature of the higher band  of $\rho^T_{\bf k}$ can be obtained by using the formula $-\frac{1}{4\pi}(\partial_{k_x} {\hat{\tilde{\bm a }}}\times \partial_{k_y} {\hat{\tilde{\bm a }}})\cdot {\hat{\tilde{\bm a }}}$, where ${\hat{\tilde{\bm a }}} ={{\tilde{\bm a }}}/|{{\tilde{\bm a }}}|$. 

To determine the phase transition point, we  use the second protocol with $\varphi = \pi/2$. For the K point with momentum ${\bf k} = (4\pi/3\sqrt{3}l,0)$, the   particle density observed  becomes
\begin{eqnarray}
{N}^{\text{II}}_{\text{TOF}}({\bf k},\tau) 
\propto 1+ \cos (\Omega \tau) a_{{\bf{k}}, 1} + \sin(\Omega \tau) a_{{\bf{k}}, 3}. \label{den1}
\end{eqnarray}
For the K point, $a_{{\bf{k}}, 1}$ is very small, and thus ${N}^{\text{II}}_{\text{TOF}}$ gets a $\pi$-phase shift when $a_{{\bf{k}}, 3}$ changes sign. This is shown in Fig.\,\ref{quench}b. The point of sign change  is exactly the phase transition point.

In conclusion, we have established a link between the SPDM and the topological Hamiltonian, and propose a scheme for detecting the SPDM in experiments. This opens up the possibility to experimentally measure the Berry curvature of the topological Hamiltonian, the first Chern number, and topological phase transitions in the interacting ultracold atom systems. The scheme for measuring the SPDM proposed here can be applied to other A-B sublattice structures. Without particle-hole symmetry, only the rescaled vector $\bm{a}_{\bf k}/a_{{\bf k}, 0}$ can be obtained by fitting to the experiment. However, this rescaled vector already contains the full topological information of the system.  For very strong interaction, where the quasiparticle picture does not hold anymore, the  connection between topological Hamiltonian and the SPDM is still an open question. 
 A generalized scheme for systems with more bands (especially if more than one band is occupied) will be the subject of future research.    

\begin{acknowledgments}
Jun-Hui Zheng acknowledges useful discussions with Oleksandr Tsyplyatyev. This research was funded by the Deutsche Forschungsgemeinschaft (DFG, German Research Foundation) via Research Unit FOR 2414
under project number 277974659. This work was also supported by the DFG via the high-performance computing center LOEWE-CSC. 
\end{acknowledgments}


\begin{thebibliography}{10}
\bibitem{Hassan2010rmp}
M. Z. Hassan and C. L. Kane, Rev. Mod. Phys.  \textbf{82}, 3045 (2010).

\bibitem{Xiao2010rmp}
D. Xiao, M.-C. Chang, and Q. Niu, Rev. Mod. Phys.  \textbf{82}, 1959 (2010).

\bibitem{Gehring2013nl}
P. Gehring, H. M. Benia, Y. Weng, R. Dinnebier, C. R. Ast,
M. Burghard, and K. Kern, Nano Lett. \textbf{13}, 1179 (2013).

\bibitem{Rechtsman2013prl}
M. C. Rechtsman, Y. Plotnik, J. M. Zeuner, D. Song, Z.
Chen, A. Szameit, and M. Segev, Phys. Rev. Lett. \textbf{111}, 103901 (2013).

\bibitem{Aidelsburger2013prl}
M. Aidelsburger,  M. Atala, M. Lohse,  J.T. Barreiro,  B. Paredes, and I. Bloch, Phys. Rev. Lett. \textbf{111}, 185301 (2013).

\bibitem{Jotzu2014Nat}
G. Jotzu, M. Messer, R. Desbuquois, M. Lebrat, T. Uehlinger, D. Greif, and T. Esslinger, Nature \textbf{515}, 237 (2014).

\bibitem{Flaschner2016Sci}
N. Fl\"aschner, B. S. Rem, M. Tarnowski, D. Vogel, D.-S. L{\"u}hmann, K. Sengstock, and C. Weitenberg, Science \textbf{352}, 1091 (2016).

\bibitem{Duca2015Sci}
L. Duca, T. Li, M. Reitter, I. Bloch, M. Schleier-Smith, and U. Schneider, Science \textbf{347}, 288 (2015).

\bibitem{Ishikawa1986}
K. Ishikawa and T. Matsuyama, Z. Phys. C: Part. Field \textbf{33}, 41 (1986); Nucl. Phys. B \textbf{280}, 523 (1987).

\bibitem{Gurarie2011prb}
V. Gurarie, Phys. Rev. B \textbf{83}, 085426 (2011).

\bibitem{You2014prb}
Y.-Z. You, Z. Wang, J. Oon, and C. Xu,
Phys. Rev. B \textbf{90}, 060502(R) (2014).

\bibitem{He2016prb}
Y.-Y. He, H.-Q. Wu, Z. Y. Meng, and Z.-Y. Lu,
Phys. Rev. B \textbf{93}, 195164 (2016).

\bibitem{Wang2012prx}
Z. Wang and S.-C. Zhang, Phys. Rev. X. \textbf{2}, 031008 (2012).

\bibitem{Shindou2008}
R. Shindou and L. Balents, Phys. Rev. Lett. \textbf{97}, 216601 (2006).

\bibitem{Shindou2008prb}
R. Shindou and L. Balents, Phys. Rev. B \textbf{77}, 035110 (2008).

\bibitem{Wong2013}
C. H. Wong and R. A. Duine
Phys. Rev. A \textbf{88}, 053631 (2013).

\bibitem{Sengupta2015}
Y. Li, P. Sengupta, G. G. Batrouni, C. Miniatura, and B. Gr\'emaud
Phys. Rev. A  \textbf{ 92}, 043605 (2015).

\bibitem{Zheng2018}
J.-H. Zheng and W. Hofstetter, Phys. Rev. B \textbf{97}, 195434 (2018).

\bibitem{Abanin2012prl}
D. A. Abanin and D. A. Pesin, Phys. Rev. Lett. \textbf{109}, 066802 (2012).

\bibitem{Kumar2016prb}
P. Kumar, T. Mertz, and W. Hofstetter
Phys. Rev. B \textbf{94}, 115161 (2016).

\bibitem{Hofstetter2018}
W. Hofstetter and T. Qin, J. Phys. B: At. Mol. Opt. Phys. \textbf{51}, 082001 (2018).

\bibitem{Rachel2018}
S. Rachel, arXiv: 1804.10656 (2018).

\bibitem{Irsigler2018}
B. Irsigler, J.-H. Zheng,  and W. Hofstetter, arXiv: 1806.01598 (2018).

\bibitem{Zheng2018a}
J.-H. Zheng, T. Qin, and W. Hofstetter,
arXiv: 1805.10491 (2018).


\bibitem{Hauke2014prl}
P. Hauke, M. Lewenstein, and A. Eckardt, Phys. Rev. Lett. \textbf{113}, 045303 (2014).

\bibitem{Ardila2018ar}
L. A. P. Ardila, M. Heyl, and A. Eckardt, arXiv: 1806.0817.

\bibitem{Tarnowski2017prl}
M. Tarnowski, M. Nuske, N. Fl\"aschner, B. Rem, D. Vogel, L. Freystatzky,  K. Sengstock, L. Mathey, and C. Weitenberg,
Phys. Rev. Lett. \textbf{118}, 240403 (2017).


\bibitem{Anisimovas2015prb}
E. Anisimovas, G. \v Zlabys, B. M. Anderson, G. Juzeli\= unas, and A. Eckardt, Phys. Rev. B \textbf{91}, 245135 (2015).

\bibitem{Eckardt2017rmp}
A. Eckardt, Rev. Mod. Phys. \textbf{89}, 011004 (2017).

\bibitem{Imriska2016prb}
J. Imri\v ska, L. Wang, and M. Troyer,
Phys. Rev. B  \textbf{94}, 035109 (2016).

\bibitem{Wu2016prb}
J. Wu, J. P. L. Faye, D. S\'en\'echal, and J. Maciejko, Phys. Rev. B \textbf{93}, 075131 (2016).

\bibitem{Vanhala2016prl}
T. I. Vanhala, T. Siro, L. Liang, M. Troyer, A. Harju, and P. T\"orm\"a, Phys. Rev. Lett. \textbf{116}, 225305 (2016).

\bibitem{Rubio-Garcia2018njp}
A. Rubio-Garc\'ia and J. J. Garc\'ia-Ripoll,  New J. Phys. \textbf{20}, 043033 (2018).



\bibitem{Bloch2008rmp}
I. Bloch, J. Dalibard, and W. Zwerger,
Rev. Mod. Phys. \textbf{80}, 885 (2008).

\bibitem{Weitenberg2017ar}
M. Tarnowski, F. Nur \"Unal, N. Fl\"aschner, B. S. Rem, A. Eckardt, K. Sengstock, C. Weitenberg, arXiv: 1709.01046


\bibitem{Struck2012prl}
J. Struck, C. \"Olschl\"ager, M. Weinberg, P. Hauke, J. Simonet, A. Eckardt, M. Lewenstein, K. Sengstock, and P. Windpassinger,
Phys. Rev. Lett. \textbf{108}, 225304 (2012).

\bibitem{Chin2010rmp}
C. Chin, R. Grimm, P. Julienne, and E. Tiesinga,
Rev. Mod. Phys. \textbf{82}, 1225 (2010).


\bibitem{Niu2010rmp}
D. Xiao, M.-C. Chang, and Q. Niu,
Rev. Mod. Phys. \textbf{82}, 1959 (2010).



\end{thebibliography}
\end{document}